\documentclass[12pt]{article}
\usepackage{graphics}
\usepackage{graphicx}
\usepackage{ifthen}
\usepackage{epic}
\usepackage{eepic}
\usepackage{latexsym}
\usepackage{cite}
\usepackage{amssymb}

\title{Analysis of Scanning Tunneling Spectroscopy Experiments from First 
Principles: the Test Case of C$_{60}$ Adsorbed on Au(111) $^{**}$}

\author{\'Angel J. P\'erez-Jim\'enez,$^{*}$
 Juan J. Palacios, Enrique Louis,\\
 Emilio SanFabi\'an,
 Jos\'e A. Verg\'es}
\makeatletter
\renewcommand{\@cite}[2]{%
             {$^{[#1]}$\ifthenelse{\boolean{@tempswa}}{,#2}{}}}
\makeatother

\begin{document}
\maketitle
  
\vspace{4.5cm}
\footnoterule
\begin{itemize}
\item[[*]]
\footnotesize{Dr. A. J. P\'erez-Jim\'enez, Prof. E. SanFabi\'an\\
  Departamento de  Qu\'{\i}mica F\'{\i}sica, Universidad de Alicante\\
  San Vicente del Raspeig, Alicante E-03080 (Spain)\\
  Fax: (+34)965903537\\
  E-mail: AJ.Perez@ua.es

\noindent  Prof. J. J. Palacios, E. Louis\\
  Departamento de F\'{\i}sica Aplicada, Universidad de Alicante (Spain)

\noindent  Prof. J. A. Verg\'es\\
  Instituto de Ciencia de Materiales de Madrid, {\em CSIC} (Spain)}

\item[[**]]
\footnotesize{The authors wish to thank Drs. C. Rogero and J. I. Pascual for
  critically reading the manuscript and providing us with the experimental
  results shown in this paper. Financial support from Spanish DGICYT and 
  the Generalitat Valenciana under Grants Nos. PB96-0085, GV00-151-01 and 
  GV00-095-2 is gratefully acknowledged.}
\normalsize
\end{itemize}

\clearpage
\setlength{\baselineskip}{0.333333333in}
Many of the technological applications of fullerenes and their derivatives, 
either as lubricants, non-linear optical devices, superconductors or in
molecular nanodevices depend on the use of thin films or the interaction with
metal surfaces. This explains the tremendous effort that has been devoted by a 
number of researchers 
to understand and characterize C$_{60}$ adlayers over a wide range of surfaces,
usually by means of Scanning Tunneling Microscopy/Spectroscopy (STM/S) 
as well as with other spectroscopic techniques (See the work by Rogero 
{\em et al}\cite{JCP-116-832} and references therein).
Nevertheless STM/S data are sometimes difficult to interpret without 
aid from theoretical models. STM images and STS spectra are usually 
rationalized simplifying Bardeen's first-order formalism\cite{PRL-6-57} by 
approximating the electronic structure of the tip\cite{PRB-31-805} and that of 
the sample, as well as the tip-sample 
interaction.\cite{PRB-34-5947,PRL-81-5588,PRL-74-2102}
On the other hand, single electron tunneling effects such as the Coulomb 
blockade and Coulomb staircase are accounted for by fits to simplified 
expressions 
derived from the golden-rule.\cite{PRB-44-5919,PRB-50-8961,PRB-56-9829} Each of 
the aforementioned methods has its {\em pros} and {\em cons}, but none of them
combines the following two highly desirable characteristics: First, to
overcome Bardeen's approximation in favor of Landauer's 
formalism.\cite{Datta:book:95} This would allow a widening of the analysis 
range beyond the tunneling regime down to tip-substrate contact. Second, to be 
fully {\em ab initio}, without resorting to {\em ad hoc} parameters and 
approximations to describe the interaction between the adsorbed molecule, the 
substrate, and the tip. It is only in this way that a method can gain 
reliability and predictibility when {\em reproducing} STM/STS experiments.
We propose here a method that brings these two major premises together and
which has successfully been applied previously to the characterization
of the conductance properties in molecular nanodevices of the type metallic 
electrode + molecule + metallic 
electrode.\cite{Palacios:prb:01,Palacios:prb:02,Palacios:nano:02}
The procedure, which is termed {\em Gaussian Embedded Cluster Method} (GECM), 
computes Landauer's conductance formula through a divide-and-conquer scheme. 
This means that the electronic structure of the molecule and part of the 
electrodes is considered in full depth from first principles, since it 
embraces the most important region, while the semi-infinite non-periodic 
structure of the metal is treated with an efficient recursive model.
Herein we report its capabilities studying adsorbed Buckminster 
fullerene on a gold surface, where recent and accurate STM/S experiments
are available.\cite{JCP-116-832}

Nucleation and growth of C$_{60}$ monolayers and thin films on Au(111) surfaces 
has been widely studied.\cite{Nature-348-621,SurfSci-279-49,SurfSci-295-13,
PRB-48-18244,SurfSci-366-93,PRB-61-2263} Although binding between gold and 
C$_{60}$ is weaker than with other metals,\cite{PRB-48-18244,PRL-71-2959,
PRB-61-2263} it is far from negligible; the adsorption energy is estimated 
around 40-60 kcal/mol.\cite{SurfSci-279-49,PRB-61-2263} In fact, adsorption
of C$_{60}$ is able to lift the well-known $23\times\sqrt{3}$ Au(111) 
reconstruction, and photoemission studies of C$_{60}$ on polycrystalline 
Au\cite{PRB-44-13747,PRB-46-7873} and Au(111) surfaces\cite{PRB-61-2263}
revealed energy shifts indicative of LUMO hybridization and charge transfer
from Au to the adsorbed fullerene molecule. 
 At high coverages, closed-packed layers grow with the 
thermodynamically most stable adsorbate phase being a 
($2\sqrt{3}\times2\sqrt{3}$) $R 30^{o}$ structure with a nearly perfect 
lattice matching in which all the molecules are in equivalent surface 
sites.\cite{SurfSci-279-49} Apart from this, another superstructure forms with
crystallographic directions matching those of the substrate, resulting in 
fullerene molecules sitting on different 
adsorption sites.\cite{SurfSci-279-49,PRB-48-18244} The proposed $11\times11$
C$_{60}$ coverage\cite{SurfSci-279-49,PRB-48-18244} for this superstructure 
has been recently discovered to be composed of a smaller 2$\times$2 grid by 
Rogero {\em et al.}\cite{JCP-116-832}  In the following, we will focus on the 
characteristics of this superstructure as interpreted by Rogero 
{\em et al.}\cite{JCP-116-832} which serve as an excellent benchmark for our
calculations to be compared with. 

The main features that characterize the adsorption of the above-mentioned 
C$_{60}$ adlayers can be drawn from Figure 
\ref{fig:CITS}, where we show the STS data obtained by Rogero 
{\em et al.}\cite{JCP-116-832} from Current Imaging Tunneling Spectroscopy 
(CITS) measurements. This technique measures the tunneling current at each
point during a topographic scan for a range of bias voltages applied to the 
sample-tip system. The numerical differentiation of each curve provides the
corresponding conductance profile. From these curves it is possible to 
construct an image of the conductance at a given bias for each point like the
one appearing in the right panel of Figure \ref{fig:CITS}, which was obtained
at a bias of +0.6V.\cite{JCP-116-832} The curves plotted in the left panel of 
Figure \ref{fig:CITS} correspond  to the conductance profiles measured 
at the points indicated with the circle and triangle.\cite{JCP-116-832} 
As deduced from Figure \ref{fig:CITS} the differences in brightness between
adsorbed C$_{60}$ molecules in the CITS image correspond to differences in 
conductance height of the peaks appearing in the conductance profile. On the
other hand, the relatively 
weak C$_{60}$-Au(111) binding reflects on the sharp form of the 
peaks, with the energetic position of the C$_{60}$ LUMO level shifted towards 
the Fermi level by charge transfer from the substrate, with a HOMO-LUMO gap of 
2.3 eV. The two curves appearing in the left panel of Figure \ref{fig:CITS} 
are interpreted\cite{JCP-116-832} as 
corresponding to a 2$\times$2 superstructure of the adsorbed layer that places 
the C$_{60}$ molecules on two different adsorption sites of the underlying 
substrate surface (see Figures \ref{fig:CITS}, \ref{fig:cluB} and 
\ref{fig:cluO}).  According to this picture, C$_{60}$ molecules sit 
alternatively onto bridge and on-top sites, the different
interaction of the fullerene with the adsorption site being responsible 
for the two type of spectra. Henceforth, when a C$_{60}$ molecule sits on top
of a gold atom, where the interaction is supposed to be weaker, we end up with
the {\em bright} peak, which is sharper than that corresponding to the
molecule sitting on a bridge between two gold atoms, where the stronger 
interaction with the substrate would explain the increased width of the peak.
The above site-specific properties are consistent with a larger displacement
to the Fermi level of the peak corresponding to the bridge site 
with respect to that of the on-top one since the amount of charge transferred 
to the molecule is larger in the former than in the latter.

The characteristics and interpretation outlined in the preceding paragraph
have been confirmed by our calculations as can be seen after inspecting 
Figure \ref{fig:STSdata} where we plot the conductance
profile corresponding to the two geometries that characterize the 2$\times$2
C$_{60}$ superstructure. Since the interaction between the 
molecule and the surface is not very strong, the conductance peaks reveal the 
underlying positioning of the C$_{60}$ orbitals, from which we estimate
the HOMO-LUMO gap to be 2.9 eV, in good accordance with the experimental
STS results.\cite{JCP-116-832} The relative positioning and height of the
conductance maxima between the on-top and bridge geometries is also
reproduced.  The form of the peaks reflects,
as suggested by Rogero {\em et al.},\cite{JCP-116-832} the different
interaction of the molecules with the adsorption sites as deduced from the
Potential Energy Scan (PES) shown in Figure \ref{fig:Escan}. For each 
adsorption site we plot Density Functional Theory (DFT) calculations 
when a C$_{60}$ molecule approaches the Au(111) surface towards the on-top and
bridge sites, respectively, with either a six-member ring or a five-member 
ring facing the surface. The same equilibrium distance (2.75\AA) is obtained
for all the geometries which coincides with that found in C$_{60}$-gold 
nanobridges\cite{Nature-407-57} and is consistent with Altman and Colton 
suggestion of no height differences between adsorbed C$_{60}$ molecules on
Au(111).\cite{SurfSci-279-49,SurfSci-295-13} As seen from Figure 
\ref{fig:Escan} the adsorption energy is smaller when the molecule sits on top
of a gold atom (16 kcal/mol) than when the fullerene binds to a bridge site
(35 kcal/mol), which explains the narrower and higher form of the conductance
maximum corresponding to the on-top geometry with respect to the bridge site.  
Once the aforementioned values are corrected by considering the lateral 
interaction of the C$_{60}$ monolayer \cite{PRB-61-2263}, estimated from the 
Lennard-Jones potential to be of about 25 
kcal/mol\cite{PRB-61-2263}, they lie within the experimental 
margins discussed above. The stronger Au-C$_{60}$ interaction when the 
molecule sits on the bridge site is consistent with a larger
amount of charge transferred from the gold surface to the molecule: 
0.8 electron, as compared to the value of 0.5 electron 
obtained for the on-top site. This gap accounts for the different
alignment of the Fermi level, which for the bridge geometry lies closer to the 
LUMO-derived orbitals, in complete agreement with Rogero {\em et al.} 
findings. The amount of transferred charge calculated by us is in 
accordance with recent results derived from photoemission 
spectra\cite{PRB-61-2263}, which give 0.8$\pm$0.2 electrons per fullerene 
molecule adsorbed on Au(111). 
Finally, we point out the fact that there exist minor differences in binding 
energy with respect to the symmetry axis that points towards the surface.
Actually, this coincides with the fact that no predominant molecular 
orientation had been found\cite{JCP-116-832} and is also indicative of the 
relative weak C$_{60}$-Au(111) interaction, which allows a large degree of
rotational freedom on the adsorbed fullerene.

As mentioned before, our method is not only applicable to the tunneling regime
but also at tip-sample contact distances. This is reflected in Figure 
\ref{fig:Gscan} where we plot the conductance of the system for the two types 
of adsorption sites as the tip moves towards the adsorbed C$_{60}$. The two 
maxima correspond to conductance channels coming from the first three C$_{60}$ 
LUMO-derived orbitals, whose degeneracy has been partially removed due to the 
interaction with the gold surface\cite{JCP-116-832,PRB-61-2263} in two sets of 
two (broader peak) and one (sharper peak) resonances, respectively. These two 
peaks also appear at tunnel tip-surface distances in the experimental STS data 
of Rogero {\em et al.} (see Figure \ref{fig:CITS}). We can check that, as the 
tip comes in closer contact with the fullerene, their interaction alters the 
size, width and positioning of the peaks. The net result is an increase in the 
conductance to non-negligible values and a larger shift of the LUMO-derived 
C$_{60}$ orbitals towards the Fermi level due to the larger amount of charge 
transferred from the tip atoms to the fullerene as they approach each other.
This fact explains the slight closure of the HOMO-LUMO gap as the tip
approaches the molecule, which is apparent after inspecting the insets of
Figure \ref{fig:Gscan}. On the other hand, the relative shape of the peaks 
corresponding to the on-top and bridge geometries found at tunneling distances 
is maintained as the tip approaches the sample.
This still reflects the differences in binding strength commented above.
Figure \ref{fig:GmaxvsD} shows the value of the conductance maxima of the
second peak {\em vs.} tip-surface distance. We can see the 
change of slope due to the different conductance regimes, with the exponential 
decay typical of tunneling appearing beyond 13\AA, which agrees well with the
experimental value reported by Joachim {\em et al.}\cite{PRL-74-2102} of 
13.2\AA. The departure from a linear trend in the logarithmic representation 
of the conductance is related to the above-mentioned closure of the C$_{60}$ 
HOMO-LUMO gap. Structural deformation of the fullerene cage has not been
considered in our calculations since we have focused on the change from
tunneling to contact regimes, where the deformation of C$_{60}$ by the tip
is negligible.\cite{PRL-74-2102}

In summary, we have shown that the so-called GECM method is able to accurately 
reproduce STS spectra. The example of C$_{60}$ adsorbed on Au(111) represents 
just a good and difficult starting point, but the method looks promising as
a valuable tool in the interpretation of STM and STS spectra up to the contact 
regime. 

\section*{Methods}
Here we give a brief overview of our method; the interested reader is 
referenced to our earlier 
work.\cite{Palacios:prb:01,Palacios:prb:02,Palacios:nano:02} 
Our procedure aims at  a first-principles implementation of Landauer's 
conductance formula:
\begin{equation}
\mathcal{G}=\frac{2e^2}{h}{\rm Tr}[\hat\Gamma_L\hat{G}^r\hat\Gamma_R \hat{G}^a].
\end{equation}
In this equation, the operators $\hat\Gamma_L$ and $\hat\Gamma_R$ are built 
from the corresponding retarded, $\hat\Sigma^r$, and advanced 
$\hat\Sigma^a$, self-energy operators of the left (L) and right (R) electrodes 
according to:
\begin{equation}
\hat\Gamma_{L(R)}=i(\hat\Sigma^r_{L(R)}-\hat\Sigma^a_{L(R)}),
\end{equation}		
while $\hat{G^r}$ and $\hat{G^a}$ denote the retarded and advanced Green's
functions operators, respectively, of the whole system and where Tr represents 
the trace over the pertinent
orbitals. Simple as it looks, Landauer's formalism poses a challenging problem
since the calculation of $\hat{\Sigma}$ involves the electronic structure of a 
semi-infinite non-periodic metallic electrode while $\hat{G}$ must deal with 
the intricacies of the metal-molecule interaction which, in turn, largely 
depends on the type and positions of the atoms involved. 

Our approach to the problem consists of a divide-and-conquer scheme. First, we
perform a DFT calculation of the molecule {\em including part of the leads 
with the desired geometry}: see Figures \ref{fig:cluB} and \ref{fig:cluO}, 
where the two clusters representing the relevant part of the bridge and on-top 
geometries discussed above are depicted. We mention in passing that the DFT
calculations were performed by using the GAUSSIAN98\cite{Gaussian:98} code with
the B3LYP exchange-correlation functional\cite{JCP-98-1372} and the basis sets 
and Pseudopotentials of Christiansen {\em et al.}\cite{JCP-82-2664,JCP-93-6654}
This is also the case for the DFT calculations that led to the PES of 
Figure \ref{fig:Escan} where clusters similar to those in 
Figures \ref{fig:cluB} and \ref{fig:cluO} were used, but with the five tip 
atoms removed. Additional clusters with the five-fold axis perpendicular to the 
surface were also used to derive the corresponding PES of Figure 
\ref{fig:Escan} but are not shown for convenience. 

Since the Hamiltonian of the cluster representing the nanobridge, $\hat{H}$, 
is finite, its associated Green's functions are unsuitable for any current 
determination. In order to transform this finite system into 
an effectively infinite one we must include the self-energy of the electrodes:
\begin{equation}
[(\epsilon + i\delta)\hat I - \hat H - \hat\Sigma^r(\epsilon)]
\hat G^r(\epsilon)= \hat I  
\end{equation}

\noindent where

\begin{equation}
\hat\Sigma^r=\hat\Sigma^r_R + \hat\Sigma^r_L.
\end{equation}

\noindent
The added self-energy is determined through a Bethe lattice tight-binding 
model\cite{Louis:prb:77,Martin:prb:90} which is constrained to reproduce the 
electrode bulk density 
of states and to have the same Fermi energy as that of the
system on which the DFT calculation was initially performed.
The advantage of choosing a Bethe lattice resides in that the self-energies
can be  easily calculated through a well-known iteration 
procedure.\cite{Louis:prb:77,Martin:prb:90}
For each atom in the outer planes of the cluster we choose to add a branch
of the Bethe lattice in the direction of any missing bulk atom.
Once the Green's functions have been calculated we proceed to self-consistency
by recalculating the elements of the density matrix $P_{kl}$ according to:

\begin{equation}
\label{eqn:nab}
P_{kl} =-\frac{1}{\pi}\int_{-\infty}^{\epsilon_F}{\rm Im}
\left [\sum_{mn} S^{-1}_{km}G^r_{mn}(\epsilon)S^{-1}_{nl}\right ]
{\rm d}\epsilon
\end{equation}

\noindent where $\epsilon_F$ is the Fermi level fixed by ensuring neutrality
in the cluster, with $S$ and $G^r$ being the overlap and retarded Green's
function matrices evaluated in a non-orthogonal atomic basis set.
The new density matrix is used to recalculate the finite cluster Hamiltonian 
matrix, and the whole process is repeated until self-consistency is achieved.

\begin{figure}
\caption{\label{fig:CITS}}
\vspace{2cm}
\includegraphics{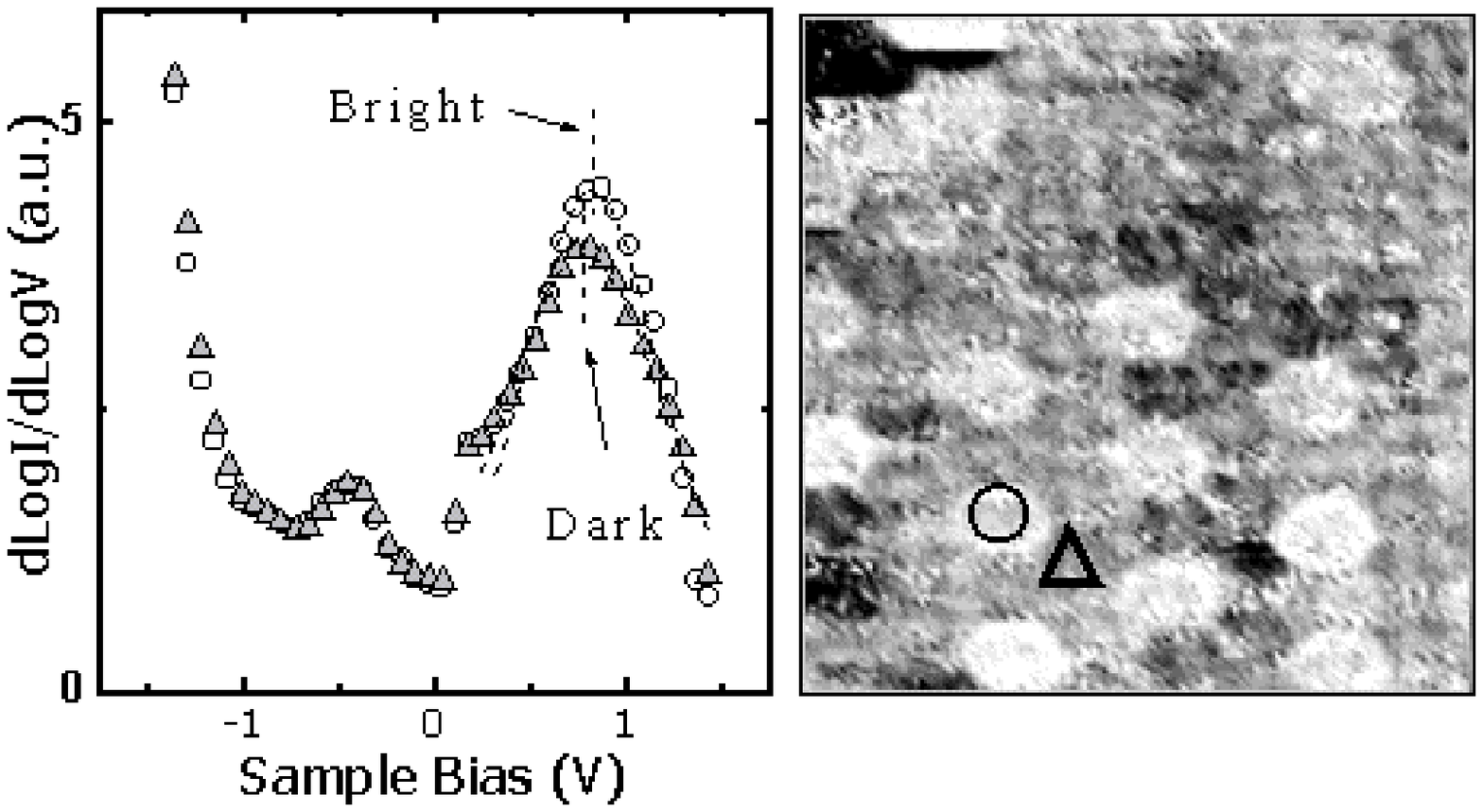}
\end{figure}

\begin{figure}
\caption{\label{fig:STSdata}}
\vspace{2cm}
\includegraphics[scale=0.9]{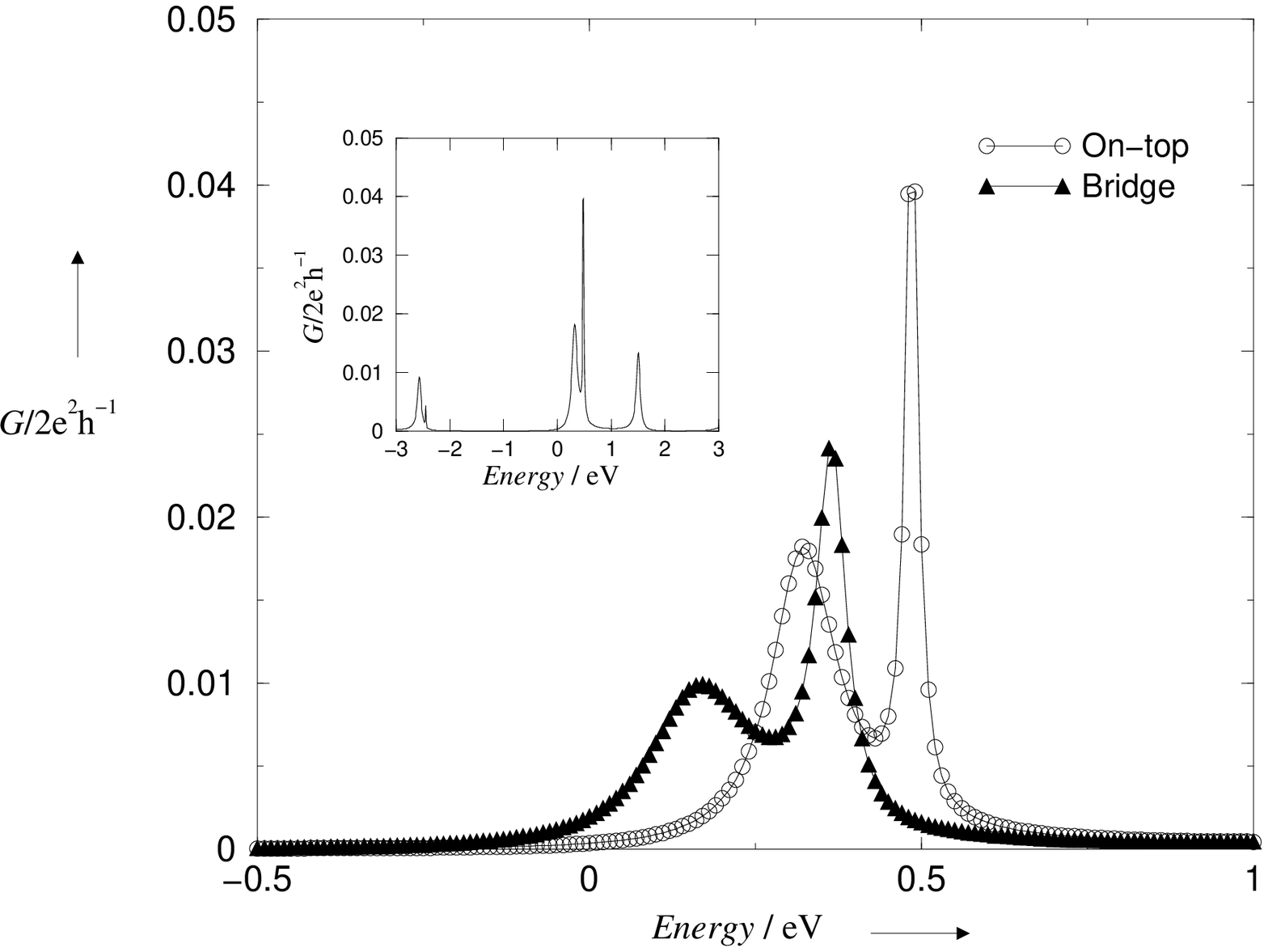}
\end{figure}

\begin{figure}
\caption{\label{fig:cluB}}
\vspace{2cm}
\includegraphics[scale=0.9]{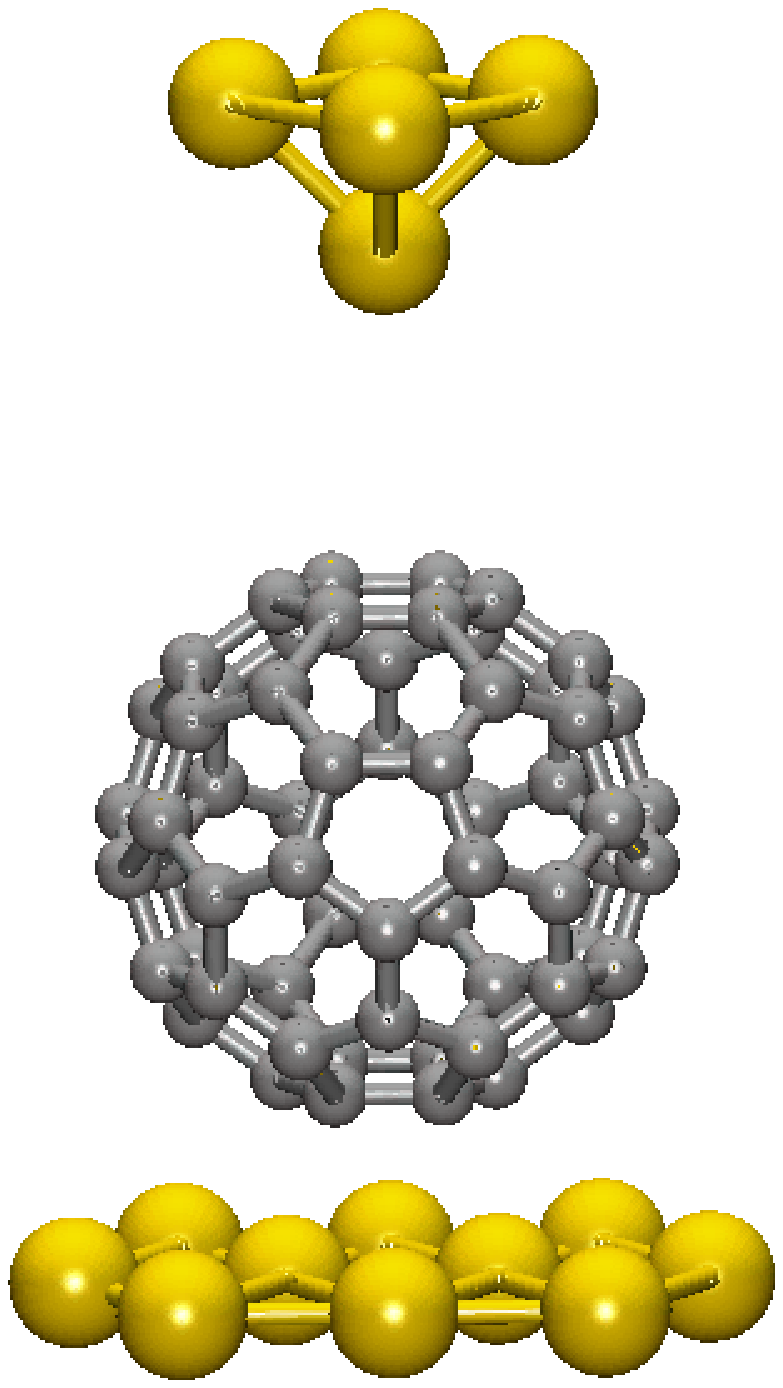}
\end{figure}

\begin{figure}
\caption{\label{fig:cluO}}
\vspace{2cm}
\includegraphics[scale=0.9]{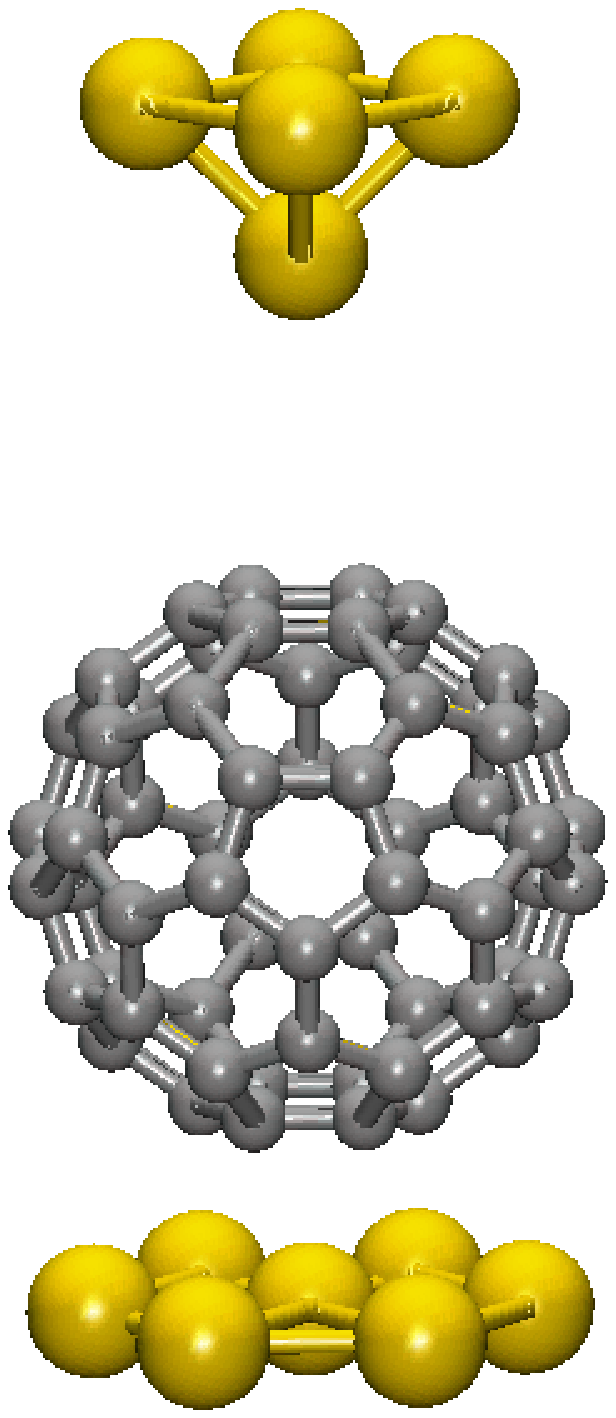}
\end{figure}

\begin{figure}
\caption{\label{fig:Escan}}
\vspace{2cm}
\includegraphics[scale=0.90]{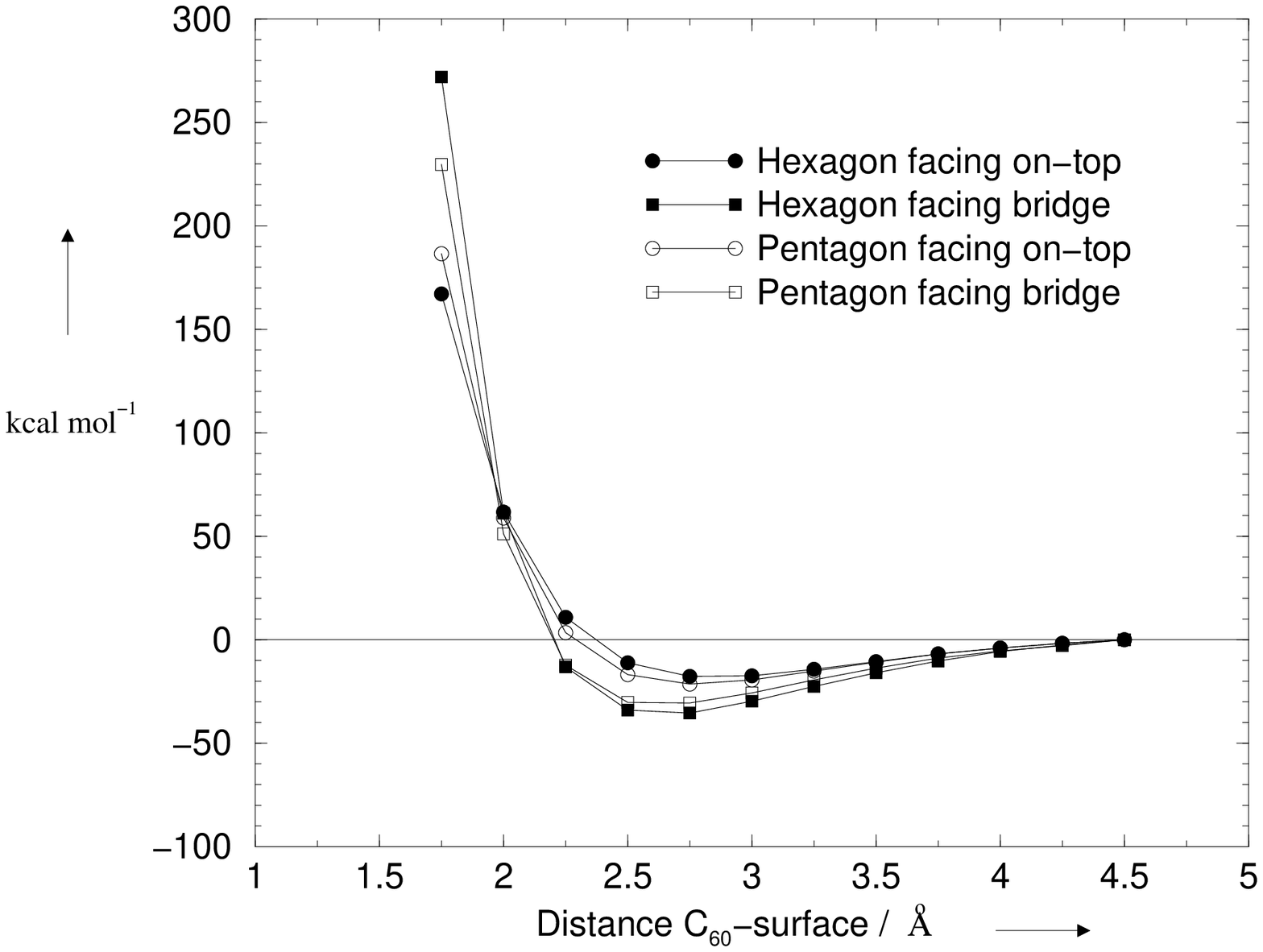}
\end{figure}

A final remark regarding the comparison between the experimental
conductance profile and the one calculated by us must be made. Strictly 
speaking only the conductance at the Fermi level should be
comparable with the experimental one, but the weak tip-C$_{60}$ 
interaction guarantees that the conductance profile will not vary strongly with
the applied voltage within the bias range used in the STM experiments. This
fact allows for a comparison between curves in the left panel of Figure 
\ref{fig:CITS} and those of Figure \ref{fig:STSdata}, although at contact
tip-fullerene distances the applied bias may produce non-trivial voltage drops.
Extension of our method to cope with generic non-equilibrium situations is 
currently being implemented.


\clearpage

\begin{figure}
\caption{\label{fig:Gscan}}
\vspace{2cm}
\includegraphics[scale=0.75]{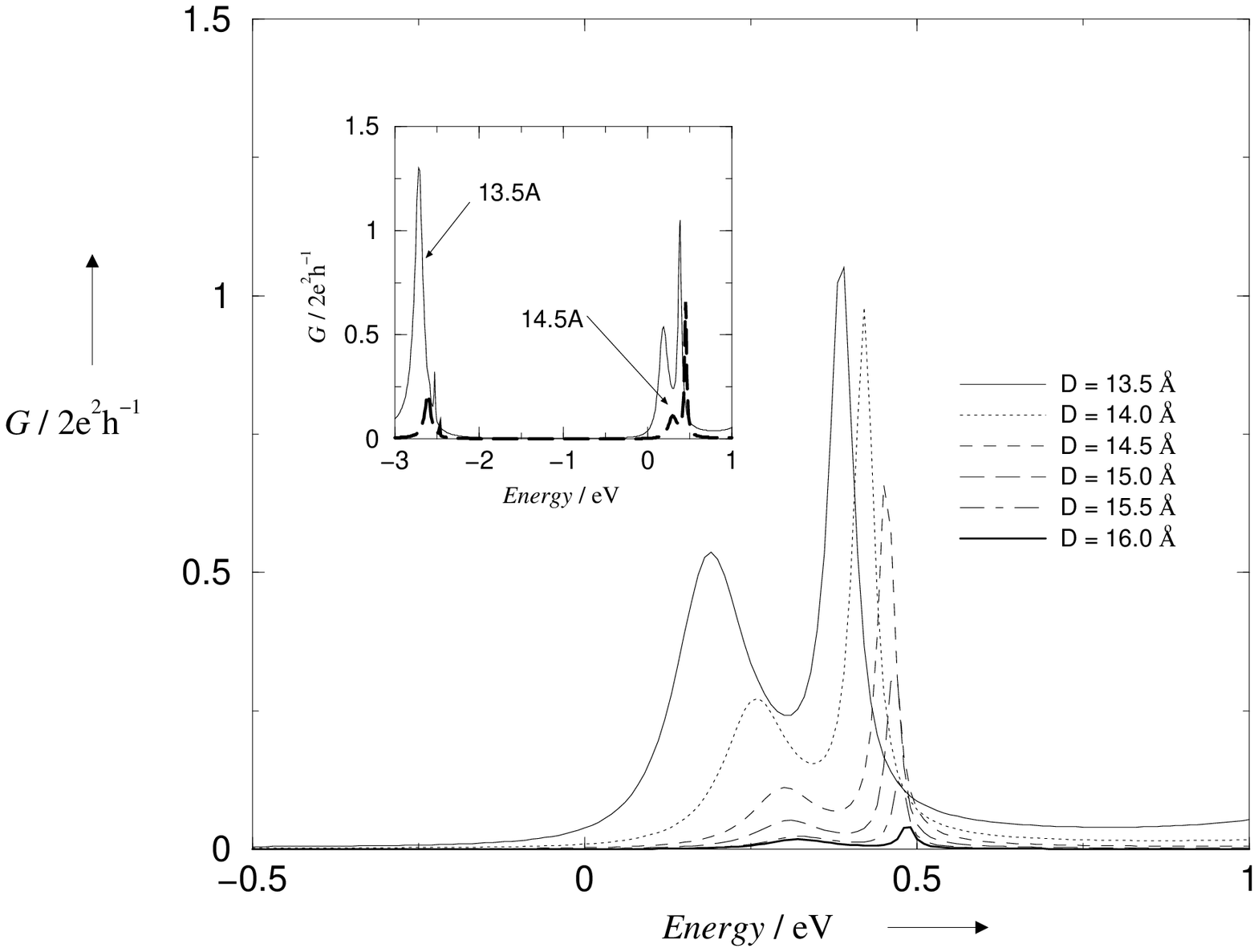}
\includegraphics[scale=0.75]{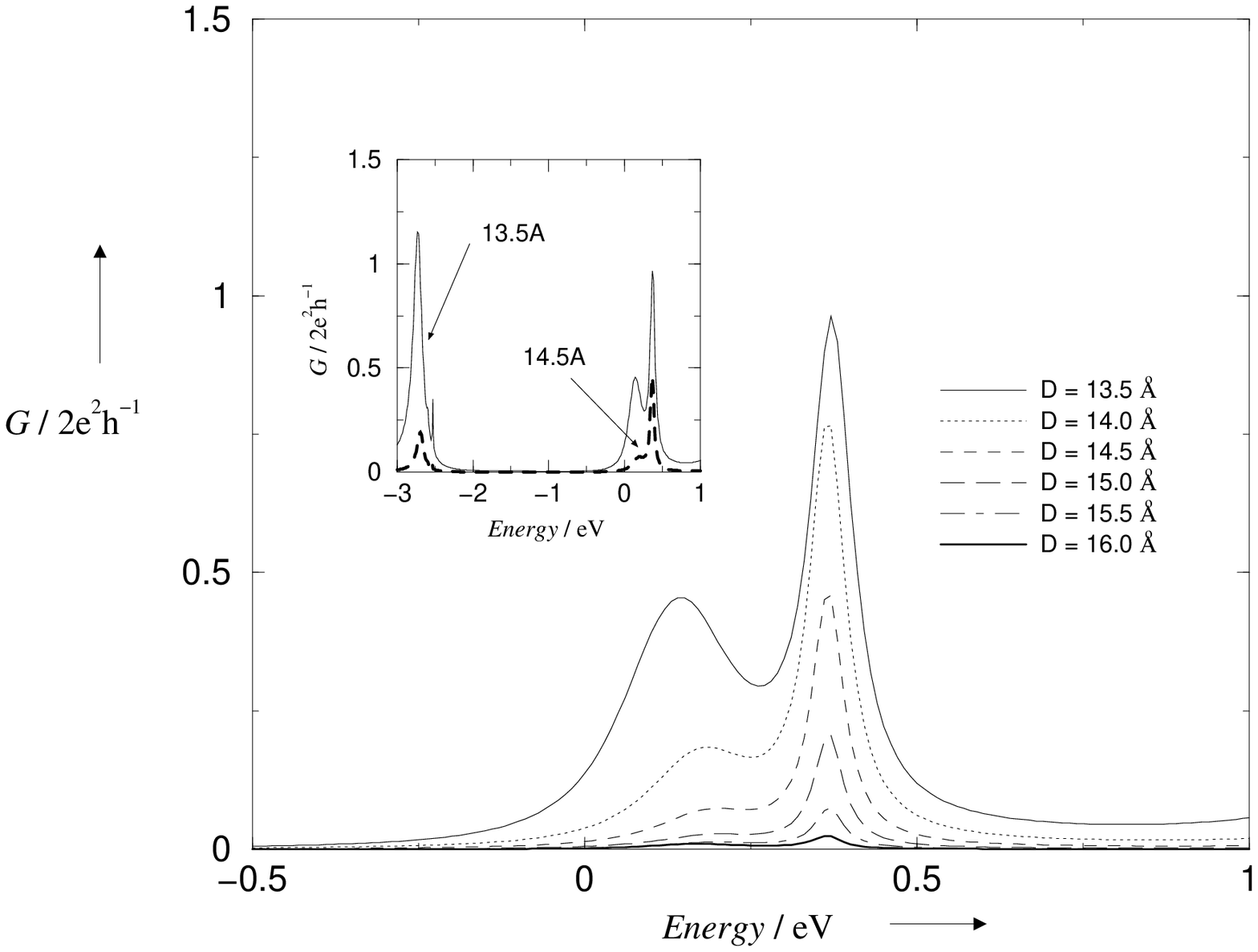}
\end{figure}

\begin{figure}
\caption{\label{fig:GmaxvsD}}
\vspace{2cm}
\includegraphics[scale=0.9]{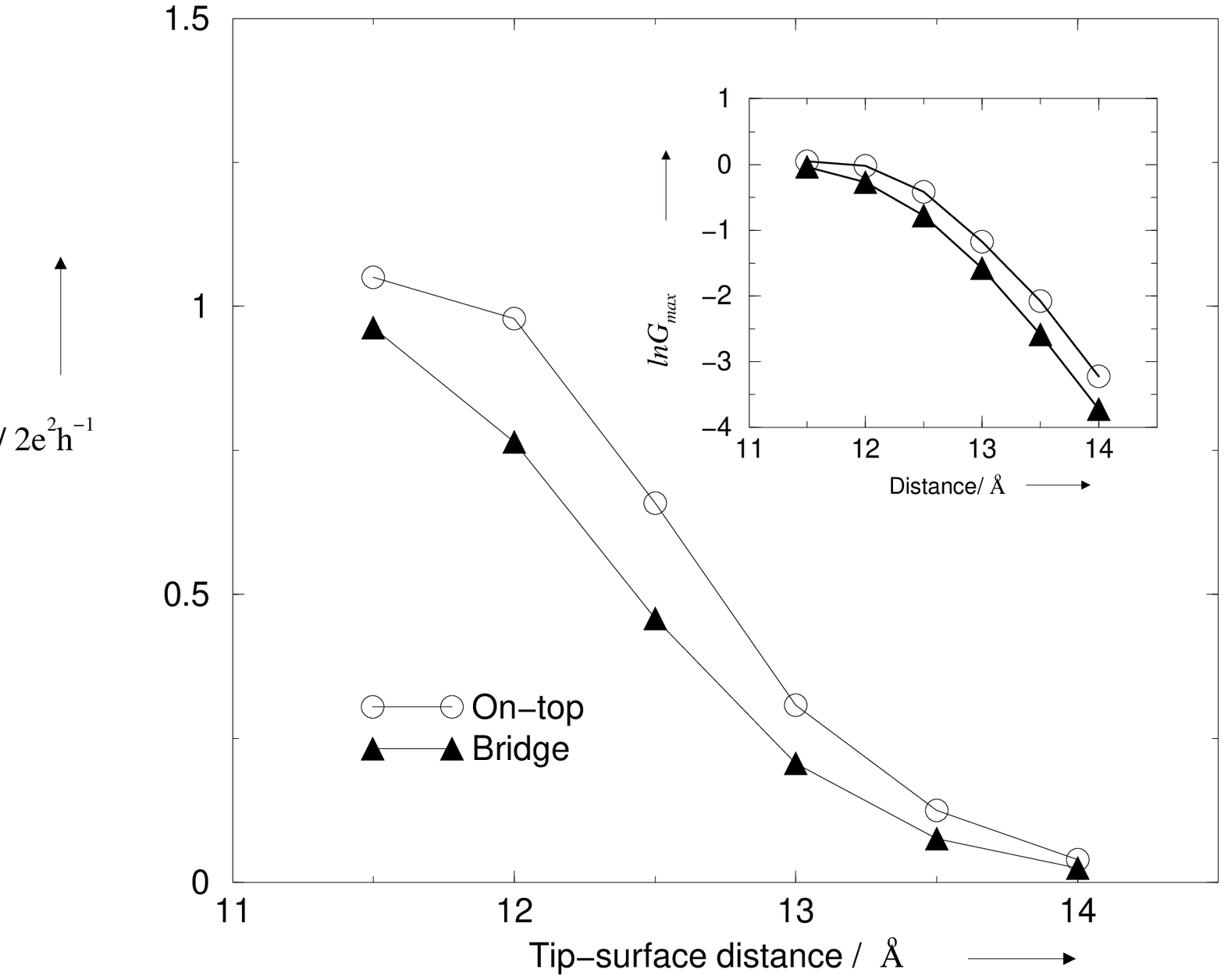}
\end{figure}

\noindent
FIG. 1: Left panel: Normalized conductance derived from CITS measurements by 
Rogero {\em et al.}\cite{JCP-116-832} for the two types of C$_{60}$ 
adsorption sites on Au(111). Right panel: CITS image by Rogero 
{\em et al.}\cite{JCP-116-832} at a bias of +0.6V, where the difference in 
brightness between on-top and bridge sites is clearly seen. The circle and 
triangle indicate the points where the curves shown in the left panel were 
obtained. The Fermi level is set to zero. The above pictures are reprints of
Figs. 4d) and 4c) appearing in {\em The Journal of Chemical
Physics}, Vol. 116, No. 2, pp. 832-836. 
\\ \\
FIG. 2: Conductance spectra calculated with our method for the on-top and 
bridge adsorption sites in the region around the Fermi level (here set to zero).
The inset shows the same data for the on-top geometry on a wider energy range, 
where the fullerene gap can be clearly appreciated. 
\\ \\
FIG. 3: Cluster model used to represent the relevant part of the 
tip-C$_{60}$-surface system. The fullerene is oriented with an hexagon
facing a bridge site of the Au(111) surface.
\\ \\
FIG. 4: Same as in Figure \ref{fig:cluB} for the on-top geometry.
\\ \\
FIG. 5: DFT Potential Energy Scan of a C$_{60}$ molecule approaching the two 
types of Au(111) surface sites (On-top and Bridge). Results with an hexagon and 
a pentagon of the fullerene facing the surface are included for completeness. 
The energy of the separated C$_{60}$ + surface is set to zero.
\\ \\
FIG. 6: Top panel: Conductance profile from tunneling to contact regime of a
C$_{60}$ molecule adsorbed on top of a gold atom in a Au(111) surface. Bottom
panel: Same as above for the C$_{60}$ molecule adsorbed on a
bridge site in a  Au(111) surface. Distance (in Angstrom) between gold surface 
and tip displayed in the legend.  The Fermi level has been set to zero in both 
sets of curves. The inset in both panels plots the same data on a wider
energy range at tip-surface distances of 13.5 and 14.5 \AA.
\\ \\
FIG 7: Maximum conductance from the second peak of Figure \ref{fig:Gscan} for
both geometries. The tunneling regime can be distinguished by the 
exponential decay in the conductance beyond 13\AA. The inset shows the same 
data with a logarithmic scale for the conductance.


\clearpage

\clearpage

\clearpage

\clearpage

\newpage
\topmargin -2cm
\textheight 23cm

\clearpage

\end{document}